# Deep Generative Models for Enhanced Vitreous OCT Imaging

Simone Sarrocco[1], Philippe C. Cattin[1], Peter M. Maloca[1,2,3], Paul Friedrich[*,1], Philippe Valmaggia[*,1,2,3]

[*]Shared last authorship

[1]Department of Biomedical Engineering, University of Basel, Allschwil, Switzerland

[2]Department of Ophthalmology, University Hospital Basel, Basel, Switzerland

[3]Moorfields Eye Hospital NHS Foundation Trust, London, EC1V 2PD, United Kingdom

**Corresponding author:** Simone Sarrocco, Department of Biomedical Engineering, University of Basel, Allschwil, Switzerland. Email: simone.sarrocco@unibas.ch

**Commercial relationships:**

| | |
|---|---|
| Simone Sarrocco: | None |
| Philippe C. Cattin: | None |
| Peter M. Maloca: | Consultant at Roche, Zeiss Forum, and holds intellectual property for machine learning at MIMO AG and VisionAI, Switzerland. |
| Paul Friedrich: | None |
| Philippe Valmaggia: | Recipient of Heidelberg Engineering and Bayer, Funding from Swiss National Science Foundation (Grant 323530_199395), Swiss Academy of Medical Sciences (YTCR 43/23) and Janggen-Pöhn Foundation. |

**Abstract**

**Purpose**: To evaluate deep learning (DL) models for enhancing vitreous optical coherence tomography (OCT) image quality and reducing acquisition time.

**Methods**: Conditional Denoising Diffusion Probabilistic Models (cDDPMs), Brownian Bridge Diffusion Models (BBDMs), U-Net, Pix2Pix, and Vector-Quantised Generative Adversarial Network (VQ-GAN) were used to generate high-quality spectral-domain (SD) vitreous OCT images. Inputs were SD ART10 images, and outputs were compared to pseudoART100 images obtained by averaging ten ART10 images per eye location. Model performance was assessed using image quality metrics and Visual Turing Tests, where ophthalmologists ranked generated images and evaluated anatomical fidelity. The best model's performance was further tested within the manually segmented vitreous on newly acquired data.

**Results**: U-Net achieved the highest Peak Signal-to-Noise Ratio (PSNR: 30.230) and Structural Similarity Index Measure (SSIM: 0.820), followed by cDDPM. For Learned Perceptual Image Patch Similarity (LPIPS), Pix2Pix (0.697) and cDDPM (0.753) performed best. In the first Visual Turing Test, cDDPM ranked highest (3.07); in the second (best model only), cDDPM achieved a 32.9% fool rate and 85.7% anatomical preservation. On newly acquired data, cDDPM generated vitreous regions more similar in PSNR to the ART100 reference than true ART1 or ART10 B-scans and achieved higher PSNR on whole images when conditioned on ART1 than ART10.

**Conclusions**: Results reveal discrepancies between quantitative metrics and clinical evaluation, highlighting the need for combined assessment. cDDPM showed strong potential for generating clinically meaningful vitreous OCT images while reducing acquisition time fourfold.

**Translational Relevance**: cDDPMs show promise for clinical integration, supporting faster, higher-quality vitreous imaging. Dataset and code will be made publicly available.

**Introduction**

Optical coherence tomography (OCT) is a non-invasive imaging technique widely used in ophthalmology to visualise different compartments of the eye at micrometre-scale resolution.[1] Despite its widespread adoption, OCT imaging faces challenges, particularly in acquiring high-quality images of the vitreous body, whose main function is to remain transparent.[2] Due to its transparency, it is not easy to properly visualise it through imaging techniques.[3] Recent work by Spaide et al. has introduced an enhanced imaging technique aimed at improving the visualisation of the vitreous body, allowing for an evaluation of anatomic changes in the vitreous associated with posterior vitreous detachment, vitreous degeneration and cisterns.[3,4] This approach involves averaging four A-scans at each position before image reconstruction, significantly boosting image quality through detailed frame averaging and high-resolution imaging.

However, obtaining high-quality OCT scans can take several minutes, significantly burdening patients and reducing clinical efficiency. Another limitation is the presence of speckle, a granular pattern inherent to OCT imaging that often obscures fine structural details.[5] Speckles arise from the interference of coherent light waves scattered by microstructures within tissue, leading to random intensity and phase variations. In OCT, this manifests as a granular, grainy texture that reduces image contrast and obscures fine anatomical details, thereby decreasing overall image quality and diagnostic clarity.[5,6] In addition, OCT images are also affected by artefacts due to patient motion during the acquisition, which results in long strips of black pixels that might completely obscure parts of the images.

There have been many different approaches to enhance the quality of OCT images. A common denoising technique is signal averaging.[7] It consists of sequentially acquiring multiple scans of the same eye location and averaging them in magnitude to increase the overall quality of the image.[8] Also, classical denoising approaches, such as median filtering, sparse and wavelet-based filtering methods, and Bayesian techniques, have shown some promise in reducing noise in OCT images.[9–11] Recent breakthroughs in machine learning have given rise to deep neural networks. Models like convolutional neural networks (CNNs) and generative adversarial networks (GANs) have been employed on medical images for many different tasks, including OCT image enhancement.[12–14]

However, each of the mentioned methods has its limitations. Traditional denoising approaches, such as filtering methods and Bayesian techniques, often blur critical anatomical details or require extensive parameter tuning, thereby limiting their clinical utility.[15] When performing signal averaging, the overall acquisition time increases linearly in $N$, creating a burden for the patient and introducing motion artefacts.[16] In multi-frame averaging, motion causes local misalignment between B-scans, especially at the edges where fewer frames overlap. This produces dark bands or black strips that partially or completely obscure affected regions. CNNs and GANs were able to generate better results thanks to the deep architecture of deep learning models.[14,17,18] However, they are still prone to unstable training, introduction of artefacts and blurriness in the output images.[19]

More recently, diffusion models, a new class of deep generative models, have been employed for many image-to-image translation tasks.[20–23] Thanks to their ability to preserve fine image details, they have been widely used in the medical field to enhance the quality of the images.[24,25] Therefore, in this work, we propose applying diffusion models for enhancing vitreous OCT images and compare their performance with other well-known deep learning models, like U-Net, Pix2Pix, and VQ-GAN.[26–28] The performance of each model is assessed from both a quantitative and qualitative perspective. For the quantitative comparison, standard image quality metrics like PSNR, MSE, SSIM, and LPIPS are used to measure the similarity between the generated images and the original high-quality OCT image; for the qualitative assessment, the model outputs are compared by visual inspection, both globally across the entire image and specifically within the vitreous region by masking out all other structures. The clinical value is assessed via two visual Turing tests performed by expert ophthalmologists from the University Hospital Basel and other clinical sites.[29,30] Additionally, a quantitative evaluation within the regions of interest inside the vitreous body is performed on newly acquired data comprised of ART1, ART10, and ART100 volumes that show the full gradation for a proper comparison. In this way, we aim to identify a deep learning-based approach capable of enhancing the quality of vitreous OCT images by significantly reducing acquisition times with respect to acquiring an ART100 image and potentially being deployed in clinical practice.

The code, together with the dataset and the visual Turing tests, will be made publicly available at https://github.com/SimoneSarrocco/vitreous_enhancement.

**Methods**

**Data Acquisition**

The main dataset used in this study consisted of retinal spectral-domain (SD) OCT images from six healthy subjects at the University Hospital Basel using a Spectralis OCT device (Heidelberg Engineering GmbH, Heidelberg, Germany) in ART10 mode. Follow-up data, used to evaluate the performance of the best model inside the vitreous body, consisted of retinal SD OCT images acquired in Lucerne from one healthy subject using the same device as the main dataset, in ART1, ART10, and ART100 modes. Ethical approval for data collection was granted from the Ethics Committee of Northwest and Central Switzerland (ID: EKNZ 2022-01091), and all participants provided informed written consent. The anonymised data were transferred under a coded data-sharing agreement to the Department of Biomedical Engineering of the University of Basel.

The OCT images of both datasets were acquired in video mode, generating consecutive B-scans of 768 × 496 pixels on a 55° field-of-view. The nominal acquisition speed was set at 20 kHz, corresponding to an integration time of 44 $\mu s$ per scan. For the main dataset, for each scan location, ten ART10 images were acquired, where each ART10 is in turn the average of ten consecutive B-scans. Since a total of 132 different locations were scanned, the dataset consisted of 1320 ART10 images, from here onwards also denoted as low-quality images or input images. For the follow-up dataset, ten ART1 volumes, ten ART10 volumes, and one ART100 volume were acquired on the same patient, where each volume consisted of 21 B-scans (i.e., 21 different eye locations).

**Weighted Signal Averaging**

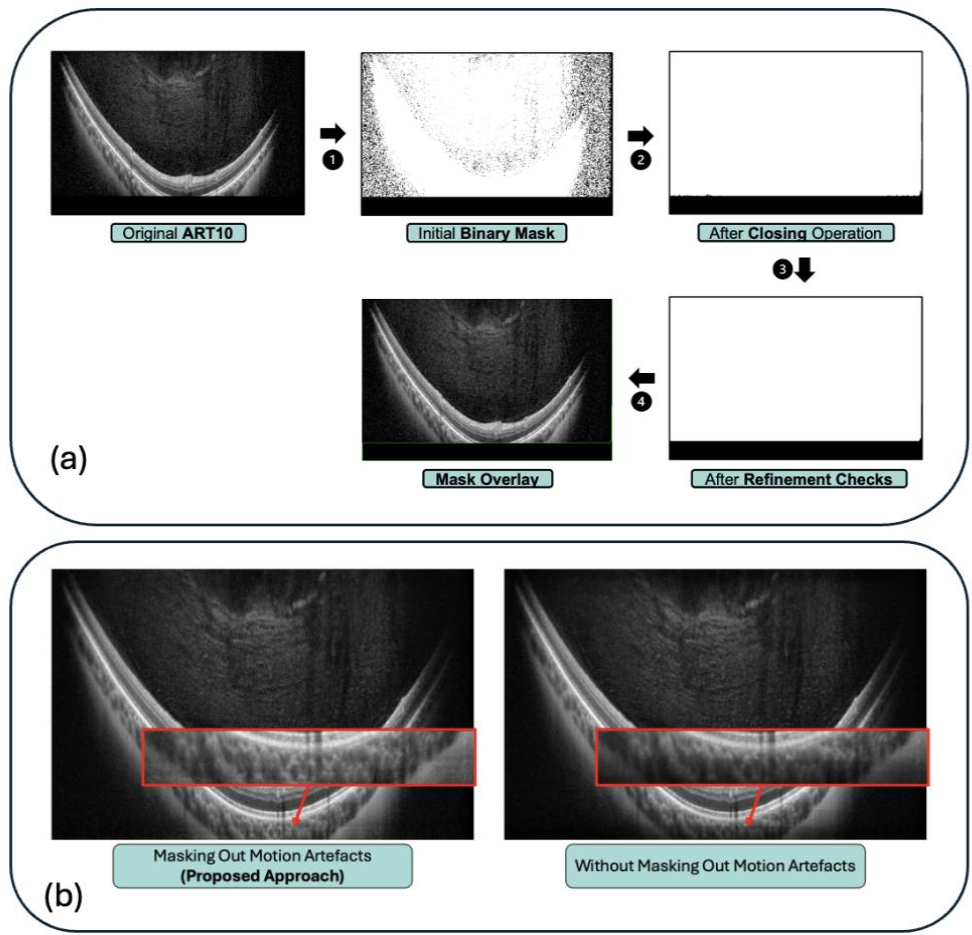

**Figure 1**. Schematic representation of the steps for generating the ground truth and visual comparison with a simple signal averaging without artefact detection. a) The steps performed to detect and mask out the artefact regions in each ART10 image are shown: 1) A binary mask of the original ART10 image is computed with a threshold of 0 to keep only black pixels. 2) The morphological closing operation is applied to fill small gaps between nearby black pixels with white pixels, thus keeping only contiguous regions of black pixels. 3) Refinement checks are applied to keep only those black pixels that are surrounded by other black pixels in at least 3 of the 4 different directions (up, down, left, right) so that small protuberances are removed from the mask. 4) The contours of the black region are drawn on the original ART10 image, showing a perfect match with the motion artefact region (long, contiguous region of only black pixels at the bottom of the image). b) Visual comparison between the proposed weighted signal averaging (left image) and simple arithmetic signal averaging (right image) for generating the ground truth. As shown in the zoomed area in the red rectangle, the proposed approach led to an increased level of detail and brightness in regions where many input ART10 images had a motion artefact.

In Figure 1a, a visualisation of the main steps of the proposed averaging method is provided. Since many input images had motion artefacts, we developed a weighted image averaging method to generate the ground truth. Motion artefacts appear as long black strips that completely obscure part of the image. To detect them in each of the ART10 images, a combination of binary thresholding and morphological operations was used to isolate contiguous strips of black pixels. Then, the final averaged image (i.e., the ground truth) was generated by assigning a weight of 0 to all pixels inside the detected artefact regions and a weight of 1 otherwise.[31] In other words, each pixel in the ground truth image is the average of all non-artefact corresponding pixels in the input images. As shown in Figure 1b, this masked averaging approach resulted in an enhanced ground truth image with less shadowing effect and a better level of contrast in the regions with many motion artefacts in the ART10 images.

Since a target image at each eye location was generated by averaging the ten ART10 of the same location, where each of them was in turn the average of ten consecutive B-scans, we referred to it as *pseudoART100*, representing the average of 100 B-scans, yet not acquired with the original ART100 settings.

**Preprocessing**

For each pair, input and target images were normalised to a pixel intensity range of [0,1] by min-max normalisation. Padding of 8 rows of black pixels, both at the top and at the bottom of each image, was applied to make the images compatible with the architecture of the models. The final pre-processed images, ready to be fed into the models, were grayscale vitreous OCT images of size 768 × 512 pixels.

**Model Architectures**

Many different deep learning models were included in the comparison, including two different kinds of diffusion models: cDDPMs and BBDMs.[20,32–34] Diffusion models aim to enhance the quality of the ART10 images by adding different levels of noise to the corresponding pseudoART100 and learning to iteratively denoise them.[20] After learning the denoising process on the pseudoART100, only the ART10 image is given as input to the model, which returns its enhanced version by iteratively removing the noise pattern learned during training.[35] The two models were compared with a U-Net, and adversarial networks like Pix2Pix and VQ-GAN.[27,28,36] A more extensive explanation of each model architecture is provided in Supplementary Material 1.

**Training Details**

Table 1. Best configuration parameters for each deep learning model

| Parameter | U-Net | cDDPM | BBDM | Pix2Pix | VQGAN |
| --- | --- | --- | --- | --- | --- |
| Image Size | 768x512 | 768x512 | 768x512 | 768x512 | 768x512 |
| Batch Size | 1 | 1 | 1 | 1 | 1 |
| Learning Rate | 2e-5 | 2e-5 | 2e-5 | 2e-4 | G: 1e-4, D: 5e-4 |
| Optimizer | AdamW | AdamW | Adam | AdamW | Adam |
| Weight Decay | 0.01 | 0.01 | 0.01 | Linearly decaying the learning rate to 0 from epoch 100 | 0.01 |
| Dropout | 0.1 | 0.1 | 0.0 | 0.0 | 0.1 |
| Best Epoch | 50 | 300 | 170 | 200 | 10 |
| Number of Channels | 128 | 128 | 128 | G: 128, D: 64 | G: 512, D:64 |
| Training Timesteps | - | 1000 | 1000 | - | - |
| Sampling Timesteps | - | 1000 | 200 | - | - |
| Variance schedule | - | Linear | Linear | - | - |
| Channel Multiplier | (0.5, 1, 1, 2, 2, 4, 4) | (0.5, 1, 1, 2, 2, 4, 4) | (1, 2, 4, 6, 8) | (1, 2, 4, 8) | (1, 2) |
| Attention Resolutions | 16, 8 | 16, 8 | 128, 64, 32, 16, 8 | - | - |
| Number of Attention Heads | 1 | 1 | 8 | - | - |
| Number of Heads per Channel | 1 | 1 | 64 | - | - |
| Number of Residual Blocks | 2 | 2 | 2 | - | 2 |
| Embedding Dimension | - | - | 8 | - | 8 |
| Number of Embedding Vectors | - | - | 16,384 | - | 16,384 |
| Loss Type | L2 | L2 | L1 | 10×L1 + LSGAN | L1 + 0.01×Adv + 0.001×Perc |
| Discriminator | - | - | - | PatchGAN | PatchGAN |
| EMA | 0.9999 | 0.9999 | 0.995 | - | - |

DDPM = Denoising Diffusion Probabilistic Model; BBDM = Brownian-Bridge Diffusion Model; VQGAN = Vector-Quantized Generative Adversarial Network; G = Generator; D = Discriminator; L1 = Mean Absolute Error; L2 = Mean Squared Error; LSGAN = Least-squared GAN; Adv = Adversarial Loss; Perc = Perceptual Loss.

To train and evaluate the models, the dataset was split into training (990 pairs of images), validation (160 pairs), and test sets (170 pairs) in a patient-wise manner to avoid having OCT scans of the same patient in more than one set. The training set comprised images from 4 patients, whereas the validation and test sets comprised images from 1 patient each. The validation set was used to choose the best setup for each model; the test set was used to compare the best-performing configurations and assess the best model from a quantitative perspective. For the BBDM, instead of using the pre-trained checkpoints of the VQ-GAN provided by the authors, which were trained on RGB images from the CelebAMask-HQ dataset, we trained our own latent space on both ART10 and pseudoART100 images from our training set to make it more suitable for reconstructing grayscale vitreous OCT images.[37] The hyperparameter choices of each best-performing setup are shown in Table 1. All models were trained and tested on a single NVIDIA® Quadro RTX 6000 24 GB (Nvidia, Santa Clara, US).

**Evaluation Metrics**

Quantitative evaluation was performed using both pixel-wise metrics like PSNR and MSE, and similarity and perceptual measures like SSIM and LPIPS, respectively.[38,39] All metrics were computed between output images and the corresponding ground truth. Moreover, since the original LPIPS is based on a CNN architecture pre-trained on ImageNet (i.e., RGB non-medical images), we decided to also use a version of LPIPS pre-trained on the RadImageNet dataset, which we called LPIPS-RAD. The latter can be found inside the MONAI Generative repository.[40,41]

To evaluate clinical applicability, relevance, and usability, expert ophthalmologists conducted two visual Turing tests and rated the models. The graders for the visual Turing tests were recruited through a request via professional ophthalmology networks of the Moorfields Eye Hospital, London, UK, and the University Eye Hospital, University of Basel, Switzerland. The first test consisted of ten questions, where in each of them a true low-quality OCT image (i.e., ART10) was shown together with the corresponding five model outputs plus the "true" high-quality image (i.e., the pseudoART100 used as ground truth in our study). In each question, ophthalmologists were asked to rank the six images from best (rank 1) to worst (rank 6) based on which they thought was the best artificially generated enhanced version of the ART10. In Figure 2a, the structure of the first visual Turing test is displayed, and it can be accessed at https://v0-visualturingtest-rho.vercel.app/.

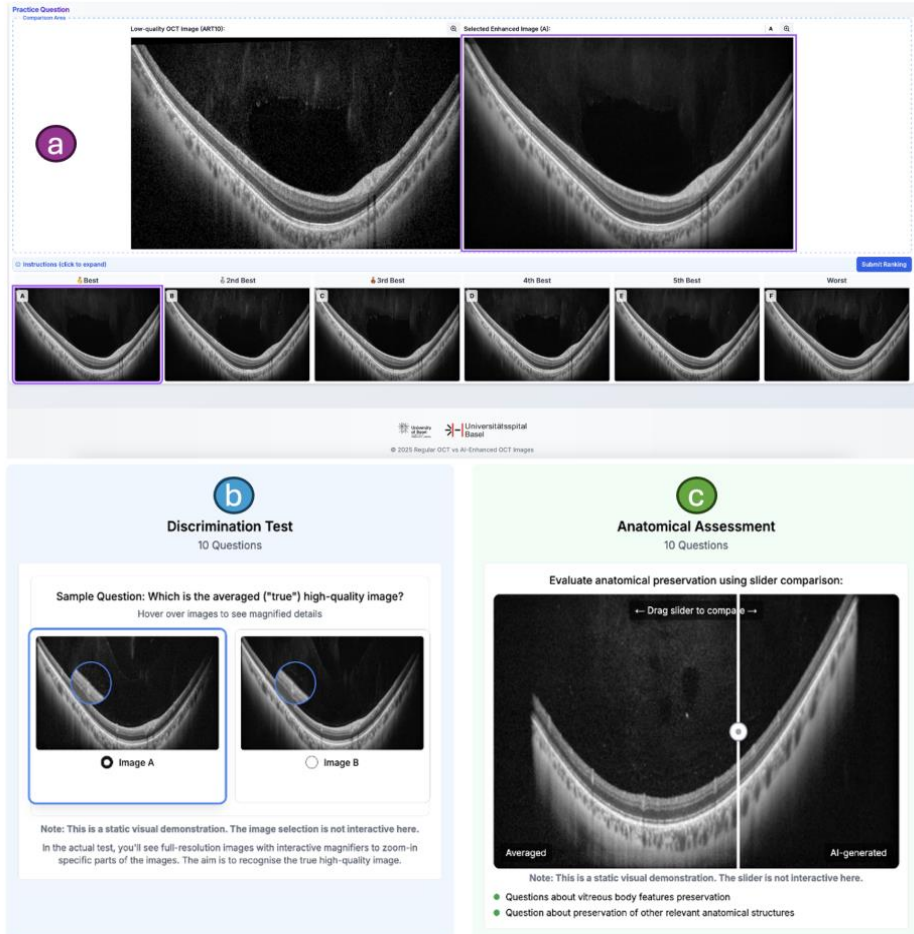

**Figure 2.** Structure of the two visual Turing tests. In Panel a (above), there is an example of a question from the first visual Turing test, where clinicians were asked to rank the 6 model outputs displayed at the bottom of the page from best (left) to worst (right). By clicking on one of the 6 images below, they could visualise it in full resolution side-by-side with the low-quality image (ART10, always displayed on the top left). In Panel b (bottom-left), a question from the first section of the second visual Turing test is shown, where clinicians had to spot the pseudoART100. In this part, they could zoom in on specific parts of both images using a synchronised magnifier that appeared on both images. In Panel c (bottom-right), a representation of the second section of the second visual Turing test is shown. Here, ophthalmologists had to compare the pseudoART100 with the generated image using a slider, to then answer all the sub-questions related to the preservation of anatomical details both in the vitreous body and in other compartments of the eye.

The second visual Turing test was performed using the output images from the deep learning model that performed best in the first test. The test was divided into two sections, each of ten questions. In the first section (Figure 2b), in each question, they had to spot the "real" high-quality OCT image, and the fool rate was computed as the percentage of wrong answers.[42] In the second section (Figure 2c), in each question, they had to answer a set of Yes/No sub-questions regarding the preservation of anatomical details of both the vitreous body and other relevant anatomical structures of the eye (Supplementary Material 2). This second test can be accessed at https://v0-new-project-atjqtgf4o4e.vercel.app/. Both Turing tests were created and deployed using the cloud platform Vercel (Vercel, San Francisco, US).

Since each eye location included ten similar ART10 images, we randomly chose one per location to prevent multiple images from the same location from appearing in the two visual tests. However, since the second test consisted of a total of twenty questions (i.e., twenty different images), for three of the eye locations of the patient included in our test set, we had to pick two images. Once the images to be included in the visual test were randomly selected, we always showed the same set of image pairs in each question to all participants.

**Results**

**Qualitative Comparison**

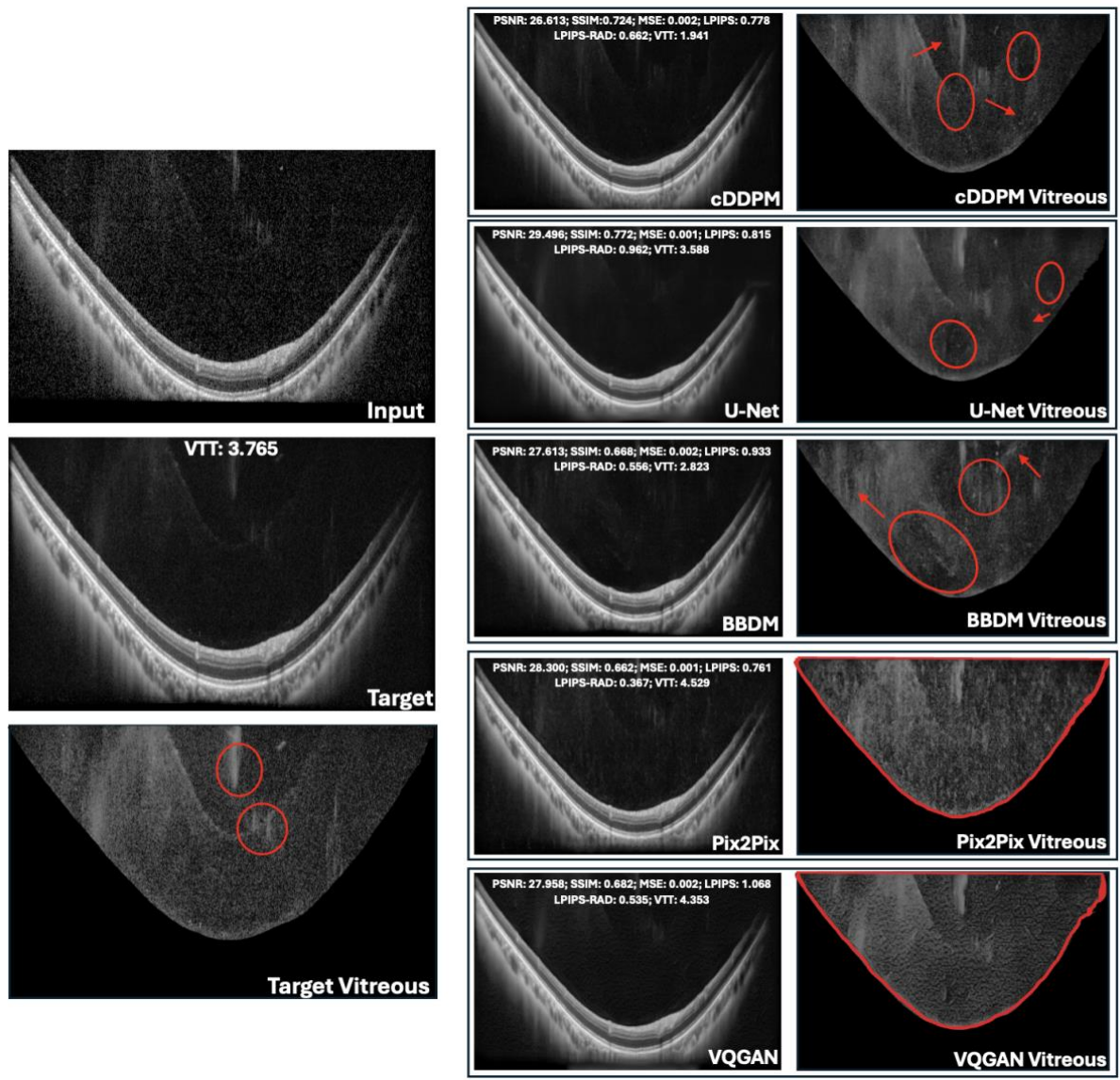

**Figure 3.** Visual comparison of model outputs and target image with an additional focus on the vitreous body. In the leftmost column, input ART10, target pseudoART100, and the segmented vitreous of the target for an example from the test set are displayed. In the two columns on the right, the model outputs from each of the five DL models and the corresponding vitreous extracted by masking out all other anatomical features in the image are shown, respectively. Before segmenting the vitreous body, histogram equalisation was applied using the open-source Fiji imaging software to enhance the contrast for a better visualisation.[44] The segmentations were obtained by using the Segment Anything demo, and they might slightly differ at the edges.[51,52] In red, some artefacts for each image are highlighted with circles and arrows. For Pix2Pix and VQGAN, the whole region is highlighted because of the presence of too many artefacts. Additionally, at the top of each generated image, the scores of all quantitative metrics with respect to the ground truth and the average rank score from the first visual Turing test are displayed. PSNR = Peak Signal-to-Noise Ratio; SSIM = Structural Similarity Index Measure; MSE = Mean Squared Error; LPIPS = Learned Perceptual Image Patch Similarity; LPIPS-RAD = LPIPS pre-trained on RadImageNet; VTT = average ranking from the first Visual Turing Test.

Figure 3 displays an example original ART10 image with a motion artefact at the bottom that obscures the choroid and the sclera. In this region, the reconstructions in the model outputs are very different from each other. For instance, in the output image from the U-Net model, no meaningful anatomical details were reconstructed, but rather a blurry region mostly filled with pixels of average intensity. A similar behaviour can be seen in the outputs from the two GANs, where there is a sort of grid-shaped artefact in the output from Pix2Pix and a strip-shaped artefact in VQ-GAN. On the other hand, the two diffusion models were able to reconstruct the anatomical structures of the choroid in a way that resembles the ground truth. When looking at the retinal layers, in the output from the BBDM, the boundaries between each layer are less defined than in the target image. In the case of VQ-GAN and U-Net, the layers are clearly visible, but there is a general loss of detail within each layer, where the reconstruction appears to be very homogeneous and overly smoothed. When isolating the vitreous region of the images by masking out all other anatomical structures, artefacts were visible across all models. The U-Net output showed a general loss of fine details due to blurriness, along with several localised artefacts (highlighted by red circles and arrows in Figure 3). Both diffusion-based models also produced artefacts within the vitreous region, with the BBDM generating anatomical structures that differed completely from the target. The vitreous generated by the cDDPM and the U-Net appeared similar, with the U-Net output being smoother and darker, while the cDDPM showed a texture pattern more consistent with the target but containing a few distorted structures. In addition, both GANs exhibited the same repeating texture patterns across the whole vitreous as in the corresponding global images. It is worth noting that the original acquisitions were also not free of artefacts. For instance, in this case, the target image presents mostly mirroring artefacts, which generated stalactite-like hyperintense regions.

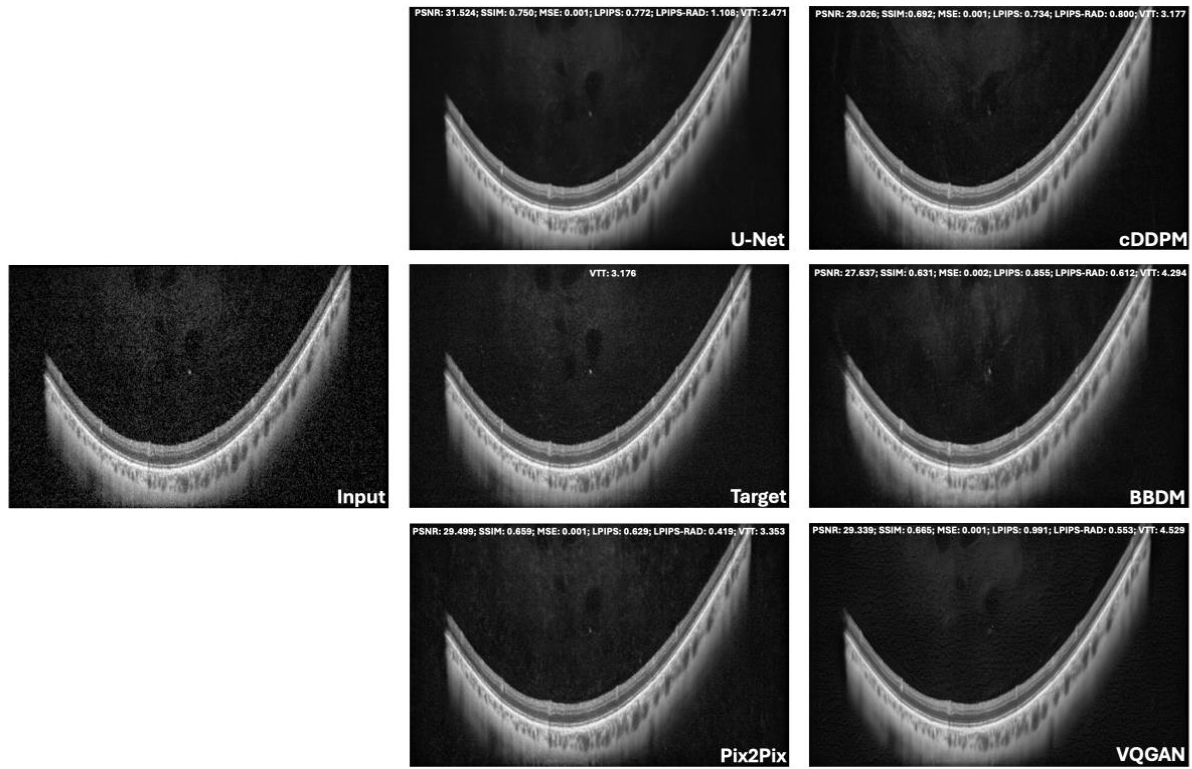

**Figure 4.** Additional visual comparison of the model outputs with the target image. On the left, the ART10 image used as input in each of the models is shown. On the two columns on the right, the output image from each of the five deep learning models, together with the target pseudoART100 image used as ground truth. Also, on top of each image, the scores of all quantitative metrics and the average ranking from the first visual Turing test are displayed. PSNR = Peak Signal-to-Noise Ratio; SSIM = Structural Similarity Index Measure; MSE = Mean Squared Error; LPIPS = Learned Perceptual Image Patch Similarity; LPIPS-RAD = LPIPS pre-trained on RadImageNet; VTT = average ranking from the first Visual Turing Test.

In Figure 4, there is an example of an input ART10 with a significant amount of speckle. It is easy to notice that in the case of Pix2Pix, the speckles are not sufficiently reduced, and the overall quality of the input ART10 image is not enhanced. Also in this case, and many cases among the test set, the output from the VQ-GAN presented a wave effect across the whole image. The retinal layers in the output images from the U-Net and BBDM were distorted with no clear boundaries. In contrast, in the reconstructed image from the cDDPM, they are well reconstructed and delineated as in the ground truth. The only exception is the retinal pigment epithelium, which is constantly brighter than the ground truth.

**Quantitative Comparison**

**Table 2.** Quantitative Metrics on the Test Set (mean ± standard deviation)

| Metrics | U-Net | cDDPM | BBDM | Pix2Pix | VQ-GAN | Input/Target |
|---|---|---|---|---|---|---|
| PSNR ↑ | **30.230 ± 2.089** | 28.615 ± 1.769 | 27.570 ± 1.326 | 28.468 ± 1.697 | 28.962 ± 1.769 | 25.336 ± 2.067 |
| SSIM ↑ | **0.820 ± 0.041** | 0.771 ± 0.043 | 0.711 ± 0.041 | 0.719 ± 0.041 | 0.748 ± 0.048 | 0.518 ± 0.072 |
| MSE ↓ | **0.001 ± 0.001** | 0.002 ± 0.001 | 0.002 ± 0.001 | 0.002 ± 0.001 | 0.001 ± 0.001 | 0.003 ± 0.002 |
| LPIPS ↓ | 0.838 ± 0.062 | 0.753 ± 0.067 | 0.870 ± 0.061 | **0.697 ± 0.058** | 0.986 ± 0.052 | 0.823 ± 0.081 |
| LPIPS-RAD ↓ | 1.069 ± 0.073 | 0.627 ± 0.107 | 0.445 ± 0.091 | **0.267 ± 0.092** | 0.552 ± 0.054 | 0.894 ± 0.098 |
| Avg. sampling time (s) | <1 | 96 | 7 | <1 | <1 | - |
| GPU Training usage | 17342 MiB | 9196 MiB | 6318 MiB | 4700 MiB | 13638 MiB | - |

The arrows indicate if higher (pointing up) or lower (pointing down) values correspond to better results. PSNR = Peak Signal-to-Noise Ratio; SSIM = Structural Similarity Index Measure; MSE = Mean Squared Error; LPIPS = Learned Perceptual Image Patch Similarity; LPIPS-RAD = LPIPS pre-trained on RadImageNet. In bold, the best scores for each metric.

The quantitative results are presented in Table 2. In the last column, the metrics between input ART10 and target pseudoART100 are computed to have baseline values for better understanding the performance of each model in generating images more similar to the ground truth than the ART10. The U-Net model achieved the best results on all pixel-based metrics, with a PSNR of 30.230 ± 2.089 and an MSE of 0.001 ± 0.001. VQGAN, cDDPM, and Pix2Pix achieved similar values of PSNR (VQGAN: 28.962 ± 1.769; cDDPM: 28.615 ± 1.769; Pix2Pix: 28.468 ± 1.697). In Figure 5, pixel-wise differences between model output and ground truth of three randomly chosen images from the test set are displayed. For each model, both the generated image and the difference map are shown, where red colours represent positive differences (i.e., pixel value of the generated image larger than the ground truth), blue colours represent negative differences (i.e. pixel value of the generated image lower than the ground truth), and white means no difference. Most of the models seemed to struggle with the reconstruction of the same anatomical structures, like the nerve fibre layer and other retinal layers in the first sample (first row of Figure 5), or the right part of the image in the second sample (second row of Figure 5). Pix2Pix, most of the time, could not reconstruct the vitreous body, as can be easily spotted from the difference map of the first sample, where the predicted pixel intensities were always higher than the ground truth.

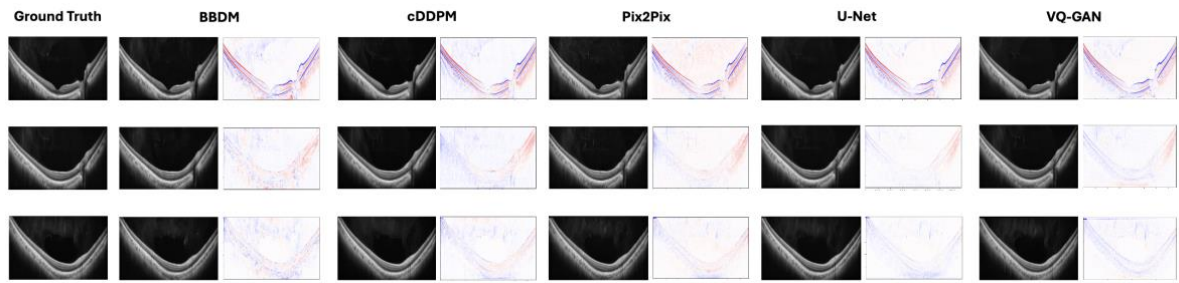

**Figure 5.** Difference maps between generated images and ground truth. In the first column, the ground truths of three samples from the test set are shown. Each of the five following columns shows the generated image (left) and the corresponding difference map (right) for one of the five deep learning models. The differences are computed as the generated image minus ground truth, where both images are in the range [0,1], leading to differences in the range [-1,1]. Red colours: positive differences; blue colours: negative differences.

In terms of SSIM, U-Net also performed best (0.820 ± 0.041). The cDDPM was the second best, with an average of 0.771 ± 0.043, followed by VQ-GAN (0.748 ± 0.048), Pix2Pix (0.719 ± 0.041), and BBDM (0.711 ± 0.041). When looking at the perceptual metrics, Pix2Pix performed best in terms of both LPIPS (0.697 ± 0.058) and LPIPS-RAD (0.267 ± 0.092). U-Net, on the other hand, generated images that were even less similar to the ground truth than the input ART10 according to both LPIPS (U-Net: 0.838 ± 0.062; Baseline: 0.823 ± 0.081) and LPIPS-RAD (U-Net: 1.069 ± 0.073; Baseline: 0.894 ± 0.098). The same happened with the scores of LPIPS for BBDM and VQ-GAN, both worse than the baseline values shown in the last column (BBDM: 0.870 ± 0.061; VQ-GAN: 0.986 ± 0.052). Among the diffusion models, cDDPM performed better than the BBDM in terms of LPIPS (0.753 ± 0.067, BBDM: 0.870 ± 0.061), but worse in terms of LPIPS-RAD (0.627 ± 0.107, BBDM: 0.445 ± 0.091).

**Visual Turing Tests**

Of the total of 399 ophthalmologists contacted, 17 participants (4.3%) registered for the first Turing test, and 7 participants (1.8%) for the second Turing test. Among the 17 participants in the first visual test, 13 had at least 5 years of experience with OCT images. In this test, the lower the average rank, and closer to 1, the better the evaluation of a model. The results are presented in Figure 6. The signal averaging approach still performed better than all deep learning models, with an average rank of 2.78. This means that, on average, clinicians still preferred the "true" high-quality image to the generated ones. Among the deep learning models, the cDDPM performed best with an average rank of 3.07, with the U-Net model right below it (3.11). Pix2Pix and BBDM were the worst, with an average rank above 4 out of 6. When performing hypothesis testing between the results of the first visual test, as shown in Figure 6, only cDDPM and U-Net were not significantly different from signal averaging, with adjusted p-values > 0.05 (Holm-corrected for multiple comparisons).[43] When stratifying the rankings by years of experience, as shown in Figure 7, the major difference was in the average ranking of the two GANs. In both sub-groups, cDDPM and U-Net were still the closest to signal averaging, with almost the same average score among clinicians with at least 5 years of experience.

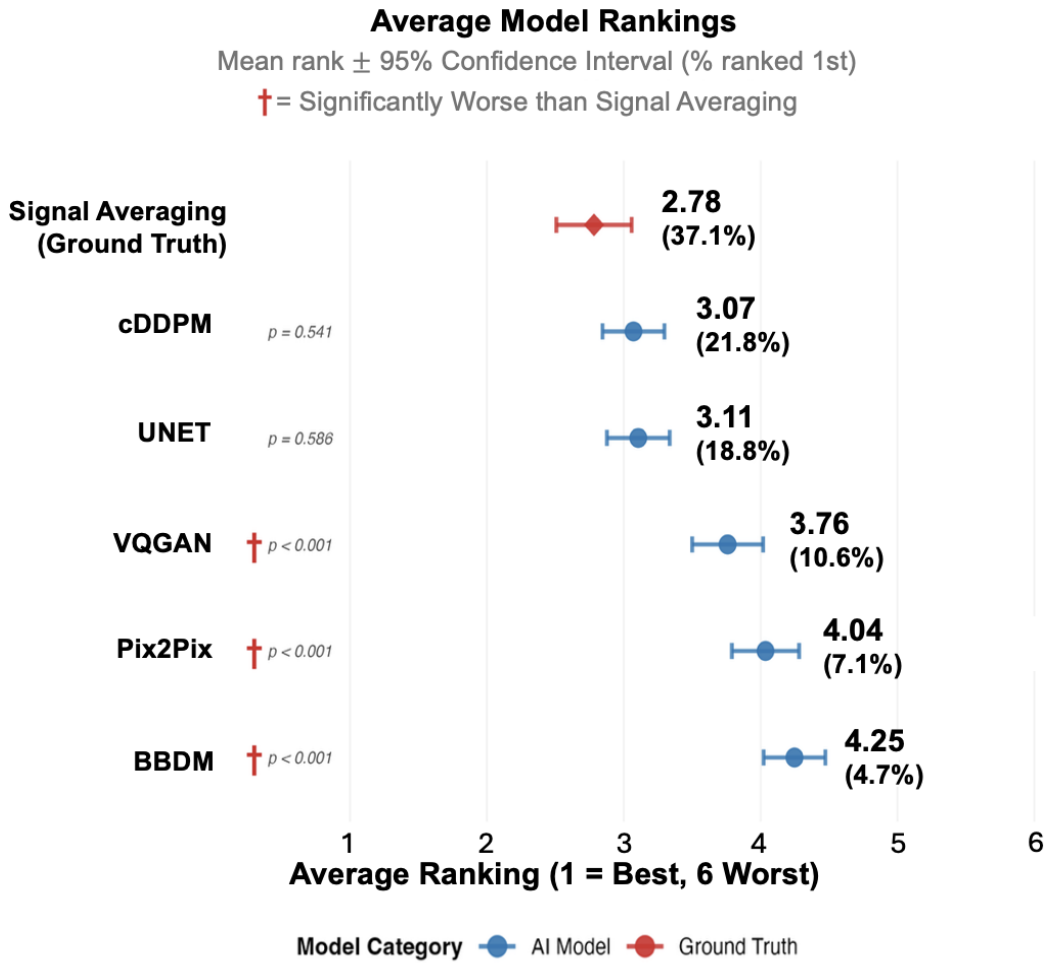

**Figure 6.** Results from the first visual Turing test. The average rankings are shown together with the 95% confidence intervals. The lower the average rank for a model, and the closer to 1, the better the performance. Models that are significantly worse than signal averaging (adjusted p-value < 0.05, Holm-corrected for multiple comparisons) are marked with a red dagger (†).

Since the cDDPM performed best among the deep learning models in the first test, the images generated with this model were used for the second test to have a more detailed clinical evaluation. Among the 7 participants in the second visual test, six had at least 5 years of experience with OCT images. The results from both sections are shown in Figure 8. In Section 1, the fool rate, defined as the percentage of how many times ophthalmologists wrongly stated that the generated image was the real high-quality OCT image, was computed. The ideal fool rate is 50%, which means that, on average, clinicians cannot distinguish between generated and real images. In our case, among the 7 participants, the average fool rate was 32.9%. In Section 2, we computed the percentage of positive answers to questions related to the preservation of the most relevant anatomical features in the images. The overall anatomical preservation of all participants was 85.7%. When focusing on the vitreous body (Figure 8a), the anatomical preservation was 78.9%, with the lowest percentage occurring in the area of Martegiani (70% of preservation), and the posterior vitreous membrane being the most preserved structure (84.3%). From the comments clinicians left when answering "No", we know that in just one case, the hyalocytes were visible in the ground truth but not in the generated image. Regarding the other anatomical compartments of the eye (Figure 8b), the retinal layers were perfectly preserved with a percentage of 100%, followed by the choroid (97.1%). The lowest preservation was about the identification of pathological structures, where the percentage was 80%. Here, we had the most heterogeneous answers, with cases in which they spotted pathological structures in 9 out of 10 images, especially dense vitreous, and cases in which they spotted none in 9 out of 10.

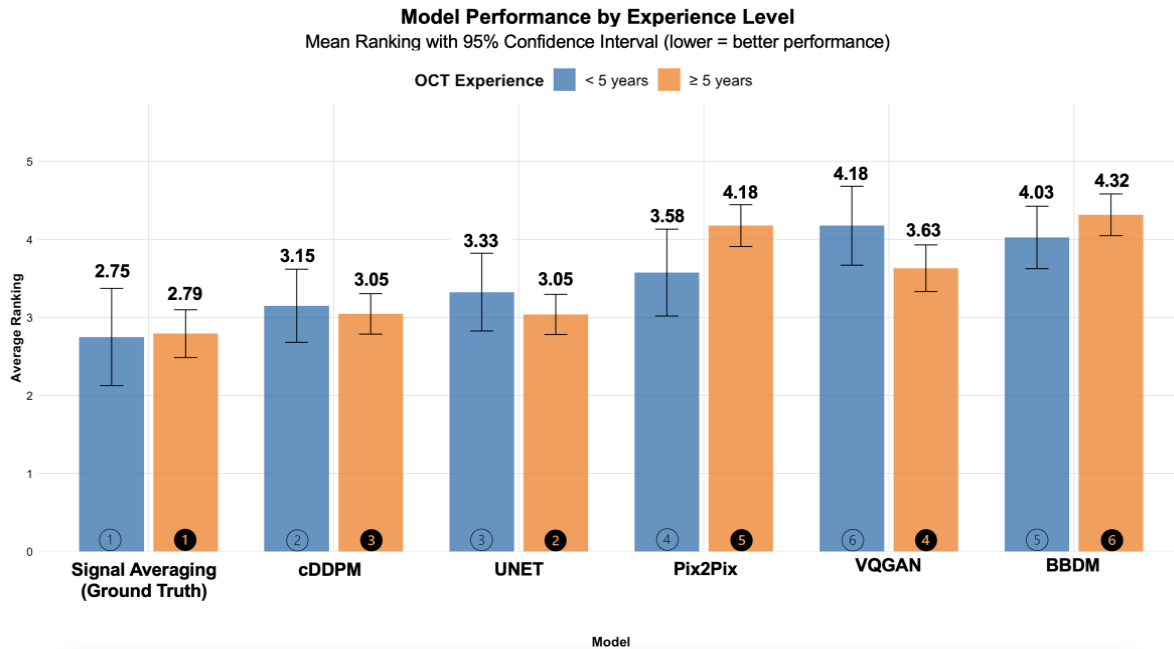

**Figure 7.** Average ranks stratified by level of experience with OCT data. In blue, the average ranks among clinicians with less than 5 years of experience are shown, whereas in orange, the corresponding scores among clinicians with at least 5 years of experience are displayed, together with the 95% confidence intervals. To help understand how each model is ranked among the two sub-groups, the ranking position is displayed at the bottom of each bar.

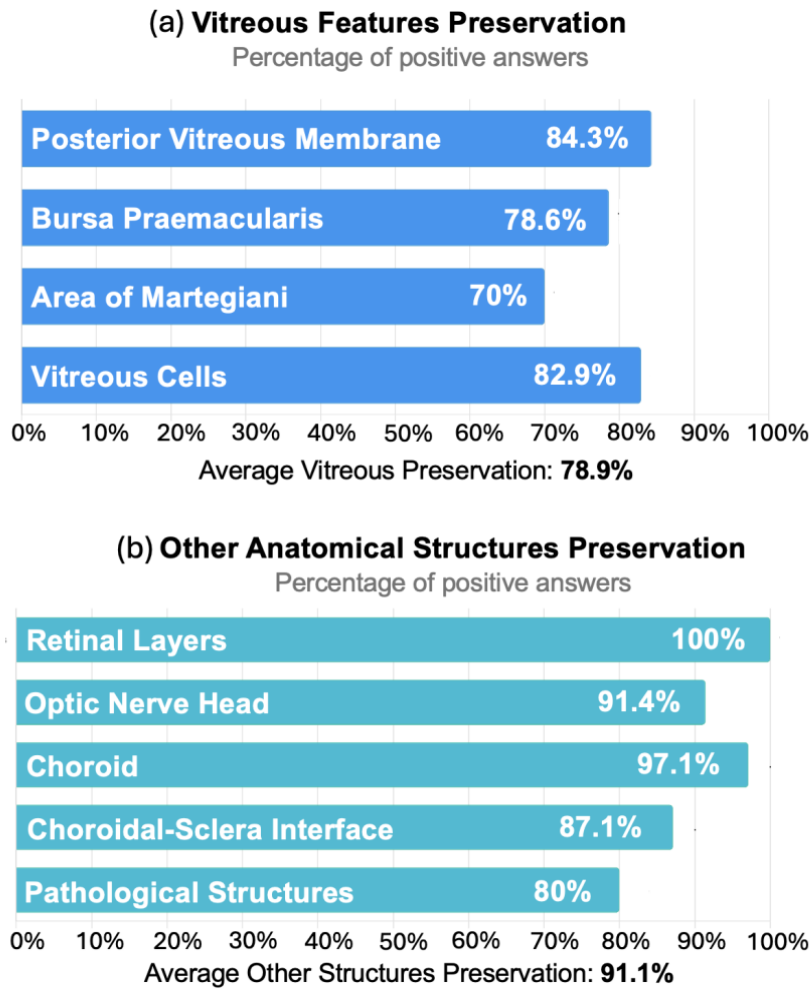

**Figure 8.** Percentage of anatomical preservation for the vitreous body and other relevant anatomical structures. In Panel a (above), the percentages of positive answers for the vitreous body are shown, which include the posterior vitreous membrane, the bursa praemacularis, the area of Martegiani, and hyalocytes. In Panel b (below), the percentage of positive answers for the remaining key anatomical structures, like retinal layers, optic nerve head, choroid, sclera, and pathological structures, is displayed.

## Quantitative Evaluation of the Best Model Inside the Vitreous Body

The quantitative performance of the cDDPM within subregions of the vitreous body was evaluated on the newly acquired data. Each ART volume comprised 21 B-scans,

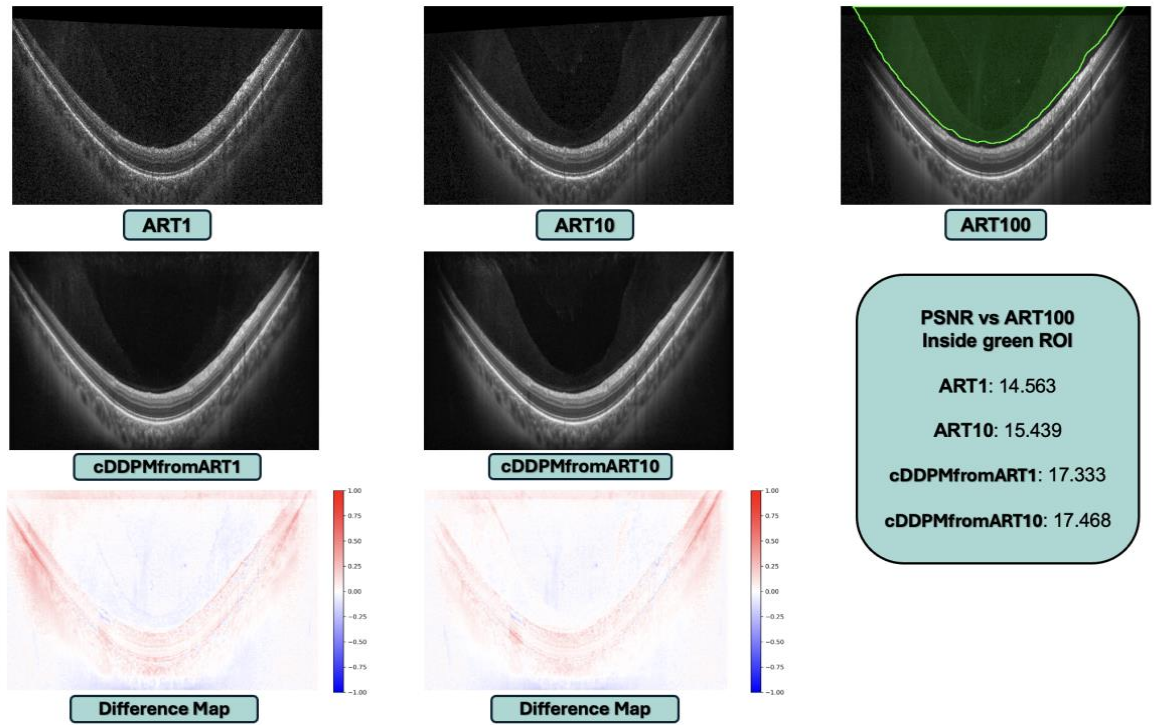

**Figure 9.** Qualitative and quantitative comparison of true ART1, ART10, and ART100 B-scans, and the corresponding images generated by the cDDPM. In the first row, the true ART1, ART10, and ART100 B-scans at the same eye location are displayed. On the ART100 image, the ROI inside the vitreous body is highlighted in green. In the second row, from left to right, the images generated from the cDDPM by using either the ART1 B-scan or the ART10 B-scan as conditioning are shown, respectively. In the last row, the corresponding difference maps, computed on a pixel level as the generated image minus the ART100 image, are displayed. Difference maps are in the range [-1,1] since they are computed between images in the pixel range [0,1].

thus 21 different eye locations. For each eye location, a region of interest (ROI) inside the vitreous body was manually segmented using the open-source Fiji imaging software.[44] The cDDPM, which was assessed as the best-performing model by ophthalmologists, was used to generate the corresponding output images starting from either a true ART1 B-scan or an ART10 B-scan. Then, the PSNR in the ROI (highlighted in green in Figure 9) was computed on a B-scan level between either true ART1 or true ART10 and the corresponding true ART100, and between generated images and the corresponding true ART100. The averaged scores across all locations

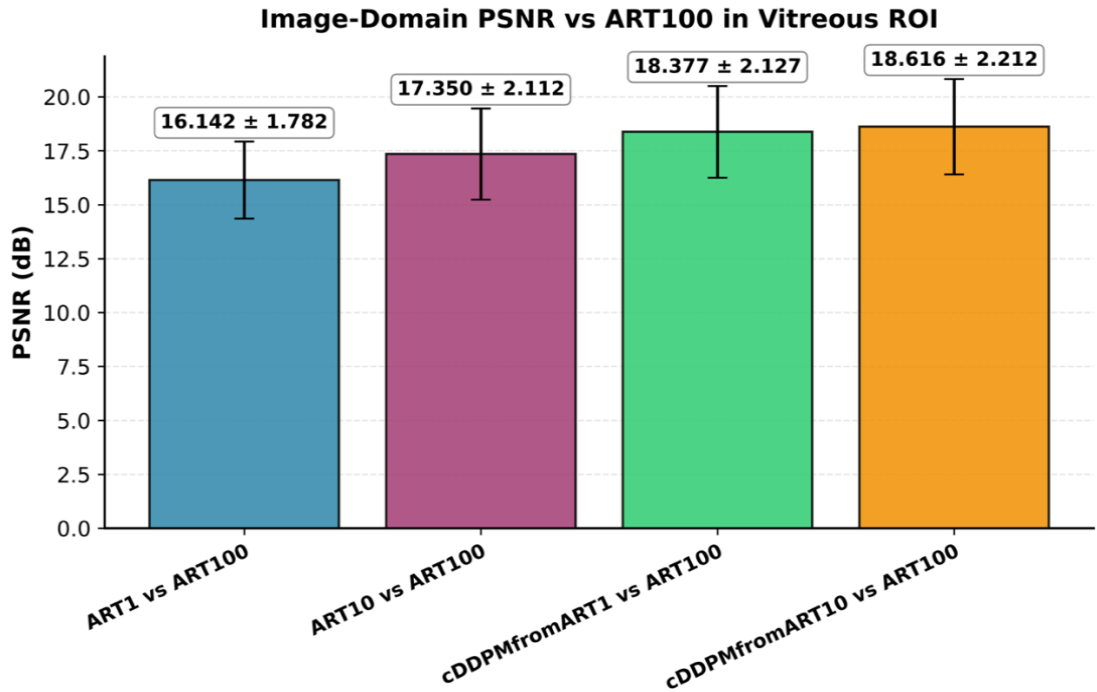

**Figure 10.** Average PSNR in the ROI inside the vitreous body across all eye locations computed on the newly acquired data. The average PSNR against the true ART100 across all eye locations is displayed for the true ART1 images, true ART10 images, and the corresponding images generated by the cDDPM starting from either true ART1 (cDDPMfromART1) or true ART10 images (cDDPMfromART10), respectively.

are shown in Figure 10. As expected, the PSNR of ART10 images (17.350 ± 2.112) is higher than the PSNR of ART1 images (16.142 ± 1.782). The images generated by the cDDPM from both ART1 and ART10 B-scans exhibit higher similarity to the ART100 reference in terms of PSNR (18.377 ± 2.127 and 18.616 ± 2.212, respectively) than the corresponding true ART1 and ART10 B-scans. In the example shown in Figure 9, the difference maps computed between each generated image and the ART100 image are very similar inside the vitreous ROI, confirmed also by the PSNR scores (17.333 for the image generated from an ART1, and 17.468 for the image generated from an ART10). Additionally, when looking at the scores in Figure 11 computed on the whole images, PSNR is even higher when using ART1 images

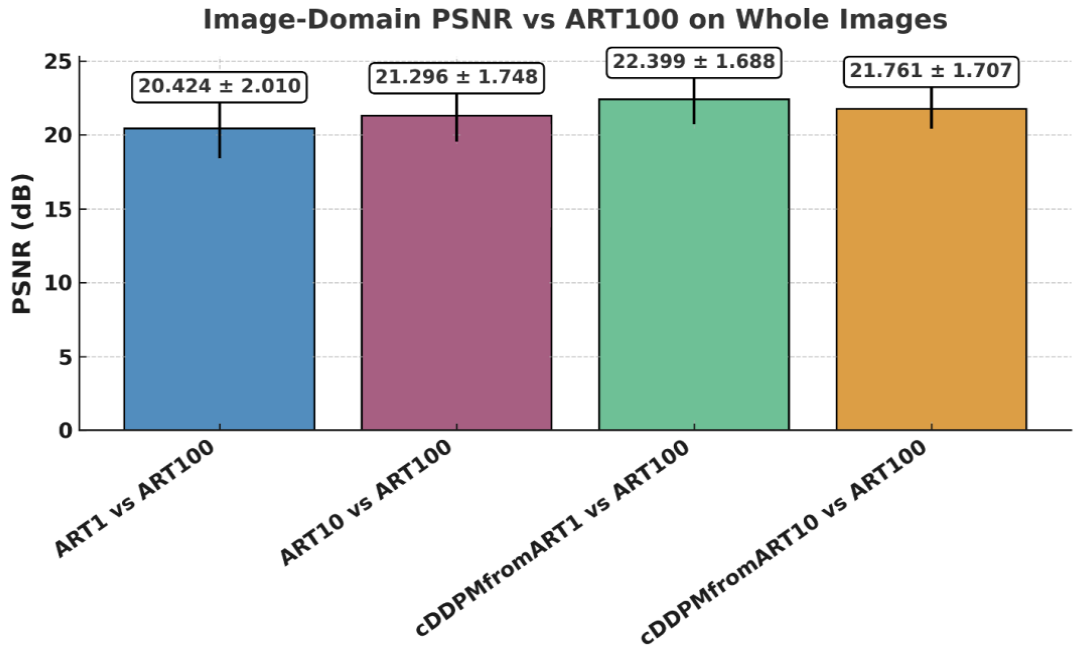

**Figure 11.** Average PSNR across all eye locations computed on the whole images of the newly acquired data. The average PSNR against the true ART100 across all eye locations is displayed for the true ART1 images, true ART10 images, and the corresponding images generated by the cDDPM starting from either true ART1 (cDDPMfromART1) or true ART10 images (cDDPMfromART10), respectively.

(22.399 ± 1.688) to generate high-quality images instead of using ART10 images (21.761 ± 1.707).

## Discussion

High image quality is essential for reliable OCT image interpretation, particularly when assessing low-contrast regions such as the vitreous body. However, acquiring such high-quality images often requires long scan times, which can be challenging for patients – especially those with poor fixation or limited cooperation – leading to incomplete or compromised acquisitions. Recent advancements in deep learning have introduced deep neural networks that address these limitations. In a supervised setting where pairs of low-quality and high-quality OCT images are available, there have been many different applications of CNNs for the denoising task.[12,13] With the rise of generative models, architectures like U-Net and GANs have been extensively explored.[17,26,27] U-Net, with its encoder-decoder structure and skip connections, has been widely used for its ability to preserve spatial resolution while reconstructing features. GANs have shown promise in generating sharper outputs through adversarial training. However, GANs are prone to unstable training and can introduce artefacts into the generated images, whereas U-Net and other CNNs can introduce blurriness, which obscures subtle anatomical features that might be critical for clinical evaluation. Diffusion models have become a valid option in medical imaging.[20–23,33] Diffusion models learn to iteratively remove noise to generate a noise-free final image. By incorporating additional input conditions to guide the generation process, cDDPMs, such as Palette, became a powerful tool for medical image-to-image translation tasks.[21,35,45]

This study showed that it is possible to use DL to enhance the visualisation of the vitreous. First, we generated high-quality images by signal-averaging them. We called these high-quality images "pseudoART100" because they resembled the quality of an ART100, which usually takes around 10 minutes to acquire. By employing the use of DL models, starting from a single ART10, we aimed at generating a clinically relevant, high-quality version as similar as possible to the corresponding pseudoART100. All DL models generated a high-quality OCT image in a much shorter amount of time than the acquisition time of an ART100. It was found that an artificial OCT image derived from the conditional diffusion model was clinically useful and could be generated, on average, in 96 seconds when sampling with the default 1000 timesteps. By considering also the acquisition time of an ART10, which is around 1 minute, we were able to generate a high-quality OCT image similar to an ART100 in approximately 2 minutes and 36 seconds, in contrast to the average 10 minutes needed to acquire an original ART100. This results in almost a 4x speedup under current experimental conditions, which could help ophthalmologists make faster acquisitions and diagnoses and make the whole process more comfortable for the patient. Interestingly, when focusing on regions inside the vitreous body, the model performed comparably well when giving as input an ART1 image instead of an ART10 image, despite having used ART10 images as conditioning during training. In other words, starting from an ART1 B-scan, on average, the model generated a vitreous region more similar to the ART100 than a true ART10, and very similar to the image generated by using the true ART10 as input. These findings also hold for the entire images rather than just inside the vitreous, suggesting that the model could be used to enhance the quality of ART1 images and potentially save even more time for generating a high-quality vitreous OCT image.

Considering both qualitative and quantitative results on the whole images, no single model clearly dominates, highlighting the complementary strengths each approach brings to vitreous OCT image enhancement. As shown in Table 2, even among the quantitative results, there are many inconsistencies depending on which metric was used to assess the DL performance. In terms of pixel-wise differences, the U-Net achieved the best results, while GANs performed the worst. When looking at the perceptual quality, we have almost the opposite situation with Pix2Pix and VQ-GAN achieving better scores than U-Net and diffusion models in terms of LPIPS-RAD. However, even among perceptual quality metrics, there is high variability in the results. For instance, depending on whether we look at LPIPS-RAD or LPIPS, BBDM is either the second-best model or the second worst, and VQ-GAN is either the third-best model in LPIPS-RAD or even worse than the baseline LPIPS between ART10 and

pseudoART100. In other words, in terms of LPIPS, the output images from the VQ-GAN were less similar to the ground truth than the ART10.

The introduction of new artefacts, especially within the vitreous body, requires careful consideration. New artefacts are likely due to the limited sample size and lack of diversity, as the dataset included only a few healthy subjects. Hallucinated anatomical structures in the generated images could critically undermine the clinical reliability of the method. Such artefacts may mislead clinicians during image interpretation, potentially resulting in diagnostic errors and patient harm. This underscores the need for rigorous evaluation strategies that go beyond standard image similarity metrics. To this end, the two visual Turing tests that we developed helped us to better understand how well each model preserved the key anatomical features usually visible in a vitreous OCT image. Models such as Pix2Pix and VQ-GAN, despite achieving strong LPIPS scores – which are intended to approximate human perception of image quality – were, on average, rated poorly by clinicians. Similarly, BBDM was the second-best model according to LPIPS-RAD but was ranked worst in the visual test. Instead, even if their average score was worse than signal averaging, cDDPM and U-Net were the only two models not significantly worse than signal averaging in terms of clinical judgment.

The correlation matrix (Supplementary Material 3) between all quantitative metrics and results from the first visual Turing test confirms the fact that the pixel-based and perceptual-based quantitative metrics led to very different rankings of the models and were different from the clinical evaluation. This highlights the need for a more reliable way of evaluating the quality of generated OCT images in terms of preservation of anatomical details, which is essential for a future deployment of the method in clinical practice. Solely relying on metrics that measure pixel-wise differences or that try to mimic how humans perceive the quality of an image might not detect subtle but important differences in how specific anatomical structures were (or were not) reconstructed, as well as the introduction of new structures that are very unrealistic from a clinical point of view.

These findings are in accordance with the results in Bayhaqi et al, where they trained ResNet and U-Net models with different loss functions to enhance OCT images of pig tissues to improve the accuracy of a tissue classifier for smart laser osteotomy.[17] In their work, they also state that quantitative image similarity metrics alone are insufficient to assess the difference between the DL models, and that a clinical evaluation is needed to understand the preservation of details achieved by each model. They also experienced blurriness in the generated images when training the U-Net with the MSE as a loss function, and just a partial removal of speckles by the model trained with an adversarial network, similar to what we observed in the images generated by Pix2Pix and VQ-GAN. Furthermore, Halupka et al. trained both a CNN with the MSE loss and a GAN to enhance retinal OCT images.[46] In their work, the CNN performed best in terms of quantitative metrics, but the GAN model was preferred from a qualitative point of view by the three clinicians who rated the generated images. Also in their work, clinicians on average still preferred the high-quality images obtained by signal averaging rather than the ones generated by DL models. However, this is expected since we trained the models by using the averaged OCT images as ground truth, so the goal was to generate OCT images with a similar - but likely lower - quality much faster, and this slight reduction in the overall image quality was spotted by the ophthalmologists.

Thanks to the second visual Turing test, we were able to quantify the anatomical preservation capabilities of the cDDPM. The results show that the conditional diffusion model was able to preserve and reconstruct most of the key anatomical features of the eye. In general, the already clearly visible structures, such as retinal layers and choroid, remained clearly visible also in the generated images. Hardly visible structures on the original OCT image were not always visible on the generated image. An explanation could be that the clinicians had different levels of confidence when looking at the images for nuanced structures in the vitreous body. Still, these findings are very promising for a future deployment of these models into clinical practice, to reduce acquisition times and assist both clinicians and patients. Future work should focus on exploring these models more by trying to develop a quantitative metric that considers pixel-wise, perceptual, and anatomical differences between generated images and ground truth. Such evaluation frameworks should explicitly quantify the presence of newly introduced artefacts, as their detection is essential for ensuring the clinical safety and reliability of these methods in practice.

Some limitations in this study might be addressed with future work. First, the relatively small sample size limits the generalizability and statistical power of the findings, potentially reducing the robustness of model performance. Since the dataset consisted exclusively of healthy subjects, the models may struggle to generalise to pathological cases (e.g., vitreomacular traction syndrome, diabetic retinopathy) or to accurately reconstruct previously unseen anatomical alterations, further limiting their applicability in clinical scenarios involving disease. Second, employing signal averaging of ten ART10 images to generate our true high-quality OCT image for training instead of directly acquiring an ART100 might have reduced the denoising capabilities of the models; that is, the quality of the pseudoART100 is likely to be lower than an original ART100. Third, it could be worth exploring alternatives to the weighted signal averaging we proposed by implementing more sophisticated and accurate ways of detecting the motion artefacts in each ART10 before averaging. Fourth, our models were trained on OCT images acquired from the same device (Heidelberg OCT Spectralis) and might not generalise well when used on images from other devices. The use of k-fold cross-validation, as well as testing the models on a larger external dataset and performing a more extensive hyperparameter tuning, might also help in achieving more robust and generalisable results. Furthermore, the implementation of more recent techniques to speed up the image generation process in the diffusion model, such as the denoising diffusion implicit model (DDIM), could potentially improve the sampling speed without compromising image quality.[47] More recently, Durrer et al. employed a variance-preserving noise schedule and reconstruction losses that enabled their 3D wavelet cDDPM to generate high-quality images for healthy tissue inpainting with just two timesteps during inference, paving the way for extremely fast and accurate medical image generation.[48] Regarding the visual Turing tests, the number of participants in the second test was lower than in the first test, but still in line with some other work. Compared to typical medical online surveys, the response rates of the conducted visual Turing tests (4.3% for the first test, and 1.7% for the second test) were within the average range between 1 to 6.3% of comparable medical surveys.[49] For instance, Bayhaqi et al. let seven experts evaluate the quality of denoised OCT images of pigs obtained from both a U-Net and a ResNet model.[17] Bellemo et al. let four clinicians evaluate the clinical relevance of their model outputs for enhancing the choroid in SD-OCT images.[50] In the latter work, in addition to the fool rate, they let clinicians measure choroidal thickness, area, volume, and vascularity index from both original images and generated ones, to then compute the correlations between the measurements.[50] For future work, it would be interesting to include a similar approach to have quantitative measurements or segmentations of the most relevant anatomical features in the images. Also, letting clinicians evaluate real high-quality OCT images and AI-generated versions separately in a random order (i.e., not one after another) instead of visualising both in the same question (as it was in Section 2 of our second visual Turing test) could be an idea to let them focus only on a single image at a time and potentially provide a more detailed and independent evaluation of anatomical preservation for each image.

**Conclusions**

This study incorporated for the first time diffusion models into a comparison of deep learning models for enhancing the quality of SD vitreous OCT images. We showed that the most common quantitative metrics for evaluating the performance of each model (i.e., PSNR, SSIM, MSE, LPIPS) can generate very different results; thus, focusing on a subset of them can be misleading. Furthermore, none of the quantitative metrics strongly correlated with the evaluation given by ophthalmologists. The mismatch observed between image similarity metrics and clinical evaluations indicates that standard performance measures may not fully reflect clinical relevance or usability, lacking an evaluation of anatomical preservation that is essential for potentially deploying the method into clinical practice. In addition, the introduction of artefacts, such as hallucinated or distorted vitreous structures, or the obscuration of important anatomical features, should be carefully assessed. should be carefully assessed. If these artefacts are not properly identified, they could compromise the method's clinical applicability.

Based on the clinical evaluation, the cDDPM seems to be a promising generative model for generating clinically relevant, high-quality vitreous OCT images that preserve key anatomical features and reduce acquisition time. Specifically, under current experimental conditions, the conditional diffusion model led to almost a 4x speedup in generating a high-quality retinal OCT image, which could potentially benefit both ophthalmologists and patients. On the follow-up dataset, the cDDPM achieved higher similarity to ART100 within the vitreous region than true ART1 and ART10 B-scans. The good performance of the model when generating high-quality vitreous OCT images starting from ART1 B-scans rather than ART10 B-scans suggests that future work could further reduce acquisition time while maintaining comparable image quality.

## Acknowledgements

The researchers would like to thank all the participants who volunteered for this study.

## Statements and Declarations

**Data Availability Statement**: The datasets acquired and/or analysed during the current study will be made publicly available from the corresponding author. Any reuse of the data must appropriately acknowledge and cite this article.

**Code Availability Statement**: The code used will be made available from the corresponding author. Any reuse of the code must appropriately acknowledge and cite this article.

**Supplementary Material 1**

Denoising diffusion probabilistic models are a class of generative models based on a parameterised Markov chain, which consists of two parts: an iterative forward diffusion process $q$ and a learned reverse process $p_\theta$.[20]

The forward process for a given image $x_0$ – in our case, a pseudoART100 – is defined by the Markov chain:

$$q(x_t|x_{t-1}) = \mathcal{N}(x_t; \sqrt{1-\beta_t}x_{t-1}, \beta_t I), \qquad (1)$$

With fixed forward variances $\beta_1, \beta_2, \ldots, \beta_T$ and an identity matrix $I$. During this process, the pseudoART100 is transformed into pure Gaussian noise by injecting at each timestep, from 1 to $T$, Gaussian noise based on Equation 1. It follows that the noisy version of the pseudoART100 can be computed as

$$x_t = \sqrt{\bar{\alpha}_t}x_0 + \sqrt{1-\bar{\alpha}_t}\epsilon, \qquad (2)$$

where $\bar{\alpha}_t = \prod_i^t(1-\beta_t)$. In the reverse process $p_\theta$, a neural network $\epsilon_\theta$ is trained to reverse the forward process and iteratively predict the slightly denoised image $x_{t-1}$ from $x_t$ for $t \in \{T, \ldots, 1\}$. Following previous work, the corresponding input ART10 image, denoted as $y$, was concatenated to $x_t$ channel-wise, and the concatenation $X_t = x_t \oplus y$ was then fed into the model.[21–23,35] The reason for the concatenation was to provide additional information to help the model reconstruct an enhanced denoised vitreous OCT image, which preserved the anatomical features of the conditioning input ART10, instead of just generating a new synthetic pseudoART100 image from the learned probability distribution. The model was trained to minimise the MSE loss

$$L_{MSE} = \|\epsilon - \epsilon_\theta(X_t, t)\|^2 = \|\epsilon - \epsilon_\theta(x_t \oplus y, t)\|^2 \qquad (3)$$

between the predicted noise $\epsilon_\theta$ and the Gaussian noise $\epsilon$ that was added to $x_0$ when computing $x_t$.

During inference, at each reversed timestep from $T$ to 1, a slightly denoised version of the pseudoART100 was computed as

$$\hat{x}_{t-1} = \frac{1}{\sqrt{\alpha_t}}\left(x_t - \frac{1-\alpha_t}{\sqrt{1-\bar{\alpha}_t}}\epsilon_\theta(X_t, t)\right) + \sigma_t z, \tag{4}$$

until we got the final denoised image $\hat{x}_0$, where $z \sim \mathcal{N}(0, I)$ represents the random component of the process and $\epsilon_\theta(X_t, t)$, is the model output at timestep $t$. The model architecture is based on the DDPM for segmentation used in the MONAI Generative tutorial.[41] Since we experienced an intensity shift in the generated images when directly predicting the noise $\epsilon_\theta$, following De Vente et al., we trained the model to predict the velocity $v = \sqrt{\bar{\alpha}_t}\epsilon - \sqrt{1-\bar{\alpha}_t}x_0$.[53] Specifically, the input was still $X_t = x_t \oplus y$, and the model output at timestep $t$ was $\hat{v}_t = v_\theta(X_t, t)$.

On the other hand, the BBDM aims at learning a stochastic Brownian bridge between the starting point (pseudoART100) and the ending point (ART10) through a bidirectional diffusion process.[33,54] The main difference with respect to the cDDPM is that the diffusion process of the BBDM is performed in the latent space of a pre-trained VQ-GAN. Moreover, since the latent representation of the ART10 is the ending point of the bridge, inference starts from there instead of from pure Gaussian noise (i.e., $x_T = y$). In this way, it is not necessary to guide the denoising process by concatenating the ART10 at each timestep as in the cDDPM. In the BBDM, Equation 2 becomes:

$$x_t = (1-m_t)L(x_0) + m_t L(y) + \sqrt{\delta_t}\epsilon$$

$$= L(x_0) + [m_t(L(y) - L(x_0)) + \sqrt{\delta_t}\epsilon], \tag{5}$$

where $m_t = \frac{t}{T}$, and $\delta_t = 2(m_t - m_t^2)$ is the variance of the forward distribution. Since the diffusion process is performed in the latent space, Equation 5 involves the latent representations of both $x_0$, denoted as $L(x_0)$, and $y$, denoted as $L(y)$. The model $\epsilon_\theta$ is then trained to predict the noise by minimising the following loss function:

$$L_{BBDM} = \left\| [m_t(L(y) - L(x_0)) + \sqrt{\delta_t}\epsilon] - \epsilon_\theta(x_t, t) \right\|^2. \tag{6}$$

During inference, the process starts from the latent representation of the ART10, and it is accelerated by reducing the number of timesteps, as with denoising diffusion implicit models.[47] The final output at timestep 0 is the latent representation of the final denoised image, which is then decoded back into the image space through the decoder of the pre-trained VQ-GAN. Our implementation of the BBDM is based on the GitHub repository of the original paper.[55]

To better understand the potential of these approaches for enhancing the quality of vitreous OCT images, the two diffusion models are compared with other well-known CNNs and GANs: a residual-attention U-Net with the same architecture as the one in the improved cDDPM from OpenAI (except for the timestep embedding) is used as CNN, whereas the default architectures of Pix2Pix and VQ-GAN are employed as GANs.[56–58] Pix2Pix is a conditional GAN that learns a mapping from paired images using a U-Net generator and a PatchGAN discriminator.[28] The two components are trained in an adversarial manner: the generator learns to produce high-quality, realistic images as similar as possible to the ground truth to fool the discriminator, while the discriminator learns to distinguish between real and generated images. VQ-GAN combines a GAN architecture with vector quantisation.[36] The model learns a discrete latent representation of the denoised image, which is then mapped back to the image domain through a decoder. In our case, the training procedure for both models consists of giving as input an ART10 image and generating output images that minimise a combination of adversarial loss, mean absolute error (i.e., $L_1$) loss, and perceptual loss with respect to the ground truth. Pix2Pix was trained by using its original GitHub repository, whereas the VQ-GAN was trained using the tutorial inside the MONAI Generative repository.[41,56]

**Supplementary Material 2**

**Figure 12.** Example question from Section 2 of the second Visual Turing Test. In this section, a total of ten questions were asked, where each of them consisted of four sub-questions related to the preservation of anatomical structures inside the vitreous body (Panel a, on the left), and five sub-questions related to the preservation of other relevant anatomical structures in the eye (Panel b, on the right), such as the optic nerve disk and the retinal layers. When answering "No", there was the chance to select a sub-option "No: not present in the image" to specify whether the anatomical feature was not present in either the real or generated image. The preservation of an anatomical feature was computed as the percentage of positive answers, that is, the percentage of "Yes" answers and "No: not present in the image" answers for the corresponding sub-question.

**Supplementary Material 3**

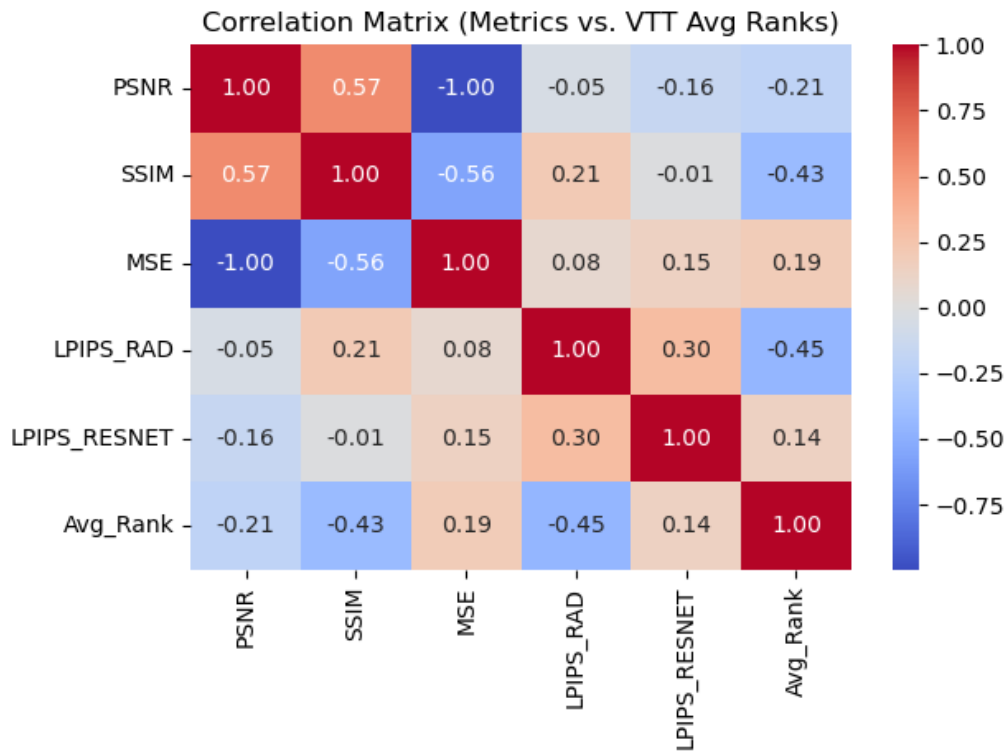

**Figure 13.** Correlation matrix between quantitative metrics and average scores from the first visual Turing test. Blue values indicate negative correlation, red values indicate positive correlation. The darker the colour, the stronger the magnitude of the correlation.